\pdfoutput=1
\documentclass[prd,twocolumn,nofootinbib]{revtex4}
\usepackage{bm} 
\usepackage{amssymb}
\usepackage{amsmath} 
\usepackage{graphicx} 
\newcommand{\E}{{\cal E}} 
\newcommand{\B}{{\cal B}} 
\newcommand{\sign}[1]{{\mbox{sgn}}(#1)}
\begin{document}
\title{Tidal interaction of a small black hole in the field of a large 
  Kerr black hole} 
\author{Simon Comeau and Eric Poisson}
\affiliation{Department of Physics, University of Guelph, Guelph, 
Ontario, Canada N1G 2W1}
\date{August 16, 2009} 
\begin{abstract} 
The rates at which the mass and angular momentum of a small black hole
change as a result of a tidal interaction with a much larger black
hole are calculated to leading order in the small mass ratio. The
small black hole is either rotating or nonrotating, and it moves on a
circular orbit in the equatorial plane of the large Kerr black
hole. The orbits are fully relativistic, and the rates are
computed to all orders in the orbital velocity $V \leq V_{\rm isco}$,
which is limited only by the size of the innermost stable circular
orbit. We show that as $V \to V_{\rm isco}$, the rates take on a
limiting value that depends only on $V_{\rm isco}$ and not on the spin
parameter of the large black hole.  
\end{abstract}  
\maketitle

\section{Introduction and summary} 

The dynamics of tidally deformed black holes has been the subject of 
vigorous investigation in the last several years \cite{alvi:01,
  poisson:04d, poisson:05, yunes-etal:06, yunes-tichy:06,
  fang-lovelace:05, taylor-poisson:08, damour-lecian:09, poisson:09, 
  johnsonmcdaniel-etal:09}.   
Most of this work was devoted to a description of the tidal
heating and torquing of a black hole within the context of a 
small-hole/slow-motion approximation, in which the calculations can be 
carried out analytically. The tidal heating of a black hole refers to
the change of its mass that occurs as a result of the tidal
interaction; tidal torquing refers to the change of its angular
momentum. These changes can be significant \cite{hughes:01, martel:04}
in realistic astrophysical situations, and the effect must be taken
into account in calculations of gravitational waves emitted by binary 
systems involving black holes. 

The small-hole/slow-motion approximation is characterized by the
condition $m/{\cal R} \ll 1$, which states that $m$, the mass of the
black hole, must be small compared with ${\cal R}$, the local radius
of curvature of the external spacetime in which the black hole
moves. For example, suppose that the black hole is moving on a
circular orbit of radius $r$ in the gravitational field of another
body of mass $M$. Then ${\cal R}^{-1}$ is of the order of the hole's
angular velocity, and we have  
\begin{equation} 
\frac{m}{\cal R} \sim \frac{m}{m+M} V^3, \qquad
V = \sqrt{\frac{m + M}{r}}, 
\end{equation} 
where $V$ is the hole's orbital velocity. One way to make this ratio 
small is to let $m/M \ll 1$; then $m/{\cal R}$ will be small
irrespective of the magnitude of $V$. This is the {\it small-hole
approximation}, which allows the small black hole to move at
relativistic speeds in the strong gravitational field of the external
body. Another way is to let $V \ll 1$; then $m/{\cal R}$ will be small
for all mass ratios. This is the {\it slow-motion approximation},
which allows the slowly-moving black hole to have a mass comparable to
(or even much larger than) $M$. 

This paper is concerned with the small-hole approximation. We take the
small black hole to move on a circular orbit in the equatorial plane
of a large Kerr black hole, and we apply the formalism developed in
Ref.~\cite{poisson:04d} to calculate the rates at which the tidal
interaction changes the mass and angular momentum of the small black
hole. The large black hole possesses a mass $M$ and an angular
momentum $J = \Xi M^2$, where $\Xi$ is the dimensionless Kerr
parameter, whose magnitude is limited by $|\Xi| \leq 1$; we adopt the 
convention that $\Xi$ is positive when the orbit of the small black
hole is corotating with the large black hole, while $\Xi$ is negative
when the orbit is counter-rotating.    

When the small black hole is nonrotating, we find that its mass and
angular momentum change according to 
\begin{subequations} 
\label{Schw} 
\begin{align} 
\frac{dm}{dv} &= \frac{32}{5} \biggl( \frac{m}{M} \biggr)^6 V^{18}
\Gamma_{\rm S}, \\
\frac{dj}{dv} &= \frac{32}{5} \sign{\Xi} \frac{m^6}{M^5} V^{15}
\Gamma_{\rm S}, 
\end{align} 
\end{subequations} 
where $v$ is advanced time, $V := \sqrt{M/r}$ an orbital-velocity 
parameter (defined in terms of the Boyer-Lindquist orbital radius
$r$), and  
\begin{equation} 
\Gamma_{\rm S} = 
\frac{(1 - 2V^2 + \Xi^2 V^4)(1 - V^2 - 2\Xi V^3 + 2\Xi^2 V^4)}
{(1 - 3V^2 + 2\Xi V^3)^2}
\label{GammaS}  
\end{equation}
is a relativistic factor that approaches unity in the slow-motion
limit $V \ll 1$. Equations (\ref{Schw}) are valid to all orders in
$V$, but are are accurate only to leading order in the small mass
ratio; they neglect terms of fractional order $m/M \ll 1$.

When the small black hole is rapidly rotating (in the sense that its
own angular velocity is much larger than the orbital angular velocity,
as discussed in Sec.~IX of Ref.~\cite{poisson:04d}), we find instead
that  
\begin{subequations} 
\label{Kerr} 
\begin{align} 
\frac{dm}{dv} &= -\frac{8}{5} \sign{\Xi} \biggl( \frac{m}{M} \biggr)^5  
\chi (1 + 3\chi^2) V^{15} \Gamma_{\rm K}, \\
\frac{dj}{dv} &= -\frac{8}{5} \frac{m^5}{M^4} 
\chi (1 + 3\chi^2) V^{12} \Gamma_{\rm K},
\end{align} 
\end{subequations} 
where $\chi := j/m^2$ is the dimensionless Kerr parameter of the small
black hole, and 
\begin{eqnarray} 
\Gamma_{\rm K} &=& 
\frac{1 - 2V^2 + \Xi^2 V^4}{(1 - 3V^2 + 2\Xi V^3)^2} 
\nonumber \\ & & \hspace*{-25pt} \mbox{} 
\times \Bigl(1 - {\textstyle \frac{4+27\chi^2}{4+12\chi^2}} V^2 
- {\textstyle \frac{4-3\chi^2}{2+6\chi^2}} \Xi V^3 
+ {\textstyle \frac{8+9\chi^2}{4+12\chi^2}} \Xi^2 V^4 \Bigr)
\label{GammaK}  
\end{eqnarray}
is the appropriate relativistic factor. It is assumed that the spin of
the small black hole is either aligned ($\chi > 0$) or anti-aligned
($\chi < 0$) with the spin of the large black hole. Equations
(\ref{Kerr}) are valid to all orders in $V$, but are are accurate only
to leading order in the small mass ratio; they neglect terms of
fractional order $m/M \ll 1$. 

The results displayed in Eqs.~(\ref{Schw}) and (\ref{Kerr}) generalize
those presented in Secs.~VIII I and IX G of Ref.~\cite{poisson:04d},
which apply when the large black hole is nonrotating ($\Xi =
0$). From Eqs.~(\ref{Schw}) we observe that when the small black hole
is nonrotating, its mass always increases, while its angular momentum
increases (decreases) when the orbit is corotating
(counter-rotating). From Eqs.~(\ref{Kerr}) we observe that when the
small black hole is rapidly rotating and its spin is aligned with that
of the large black hole, its mass decreases (increases) when the orbit
is corotating (counter-rotating), while its angular momentum always
decreases; the signs reverse when the spins are anti-aligned. 

These are the behaviors that are expected for a system in rigid
rotation. In such cases (see Sec.~IX F of Ref.~\cite{poisson:04d}) we
must have that $dm/dv = \Omega^\dagger (\Omega^\dagger 
- \Omega_H) {\cal K}$ and  $dj/dv = (\Omega^\dagger 
- \Omega_H) {\cal K}$, where $\Omega^\dagger$ is
the angular frequency of the tidal fields as measured in the moving
frame of the small black hole, $\Omega_H$ is the angular velocity of
the small black hole, and ${\cal K}$ is a quantity --- defined by
Eq.~(9.46) of Ref.~\cite{poisson:04d} --- that is constructed from the
tidal fields. According to our conventions, the signs of
$\Omega^\dagger$ and $\Omega_H$ are measured relative to the
orientation of the spin of the large black hole, so that
$\Omega^\dagger > 0$ when the orbit is corotating and $\Omega_H > 0$
when the black-hole spins are aligned. The nonrotating case
corresponds to $\Omega^\dagger \gg \Omega_H$, while the
rapidly-rotating case corresponds to $\Omega^\dagger \ll \Omega_H$. 

\begin{figure} 
\includegraphics[width=0.9\linewidth]{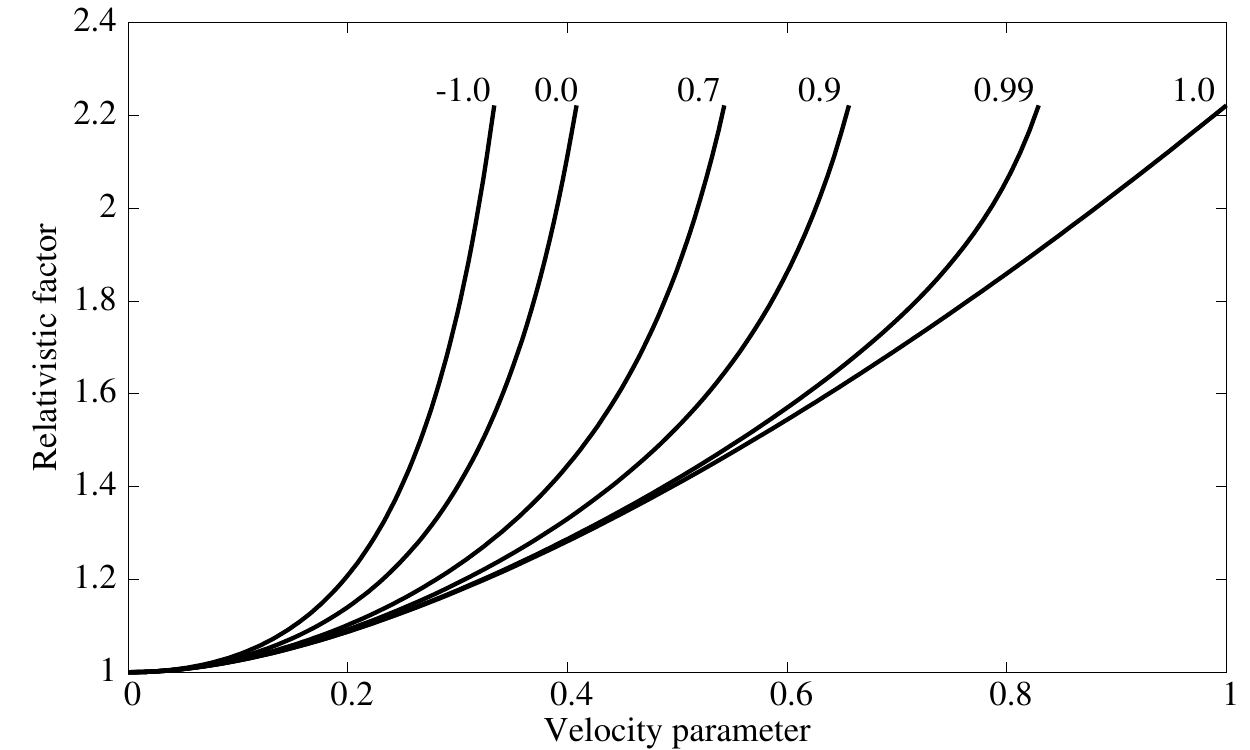}
\caption{Relativistic factor $\Gamma_{\rm S}$ for a nonrotating black
  hole, plotted as a function of the orbital-velocity parameter
  $V$. The curves are labeled by $\Xi$, which ranges from
  $\Xi = -1$ (extremal Kerr black hole, counter-rotating orbit) to 
  $\Xi = +1$ (extremal Kerr black hole, corotating orbit). The unique
  limiting value of $\Gamma_{\rm S}$ is $\frac{20}{9} \simeq 2.2222$.}    
\end{figure} 

\begin{figure} 
\includegraphics[width=0.9\linewidth]{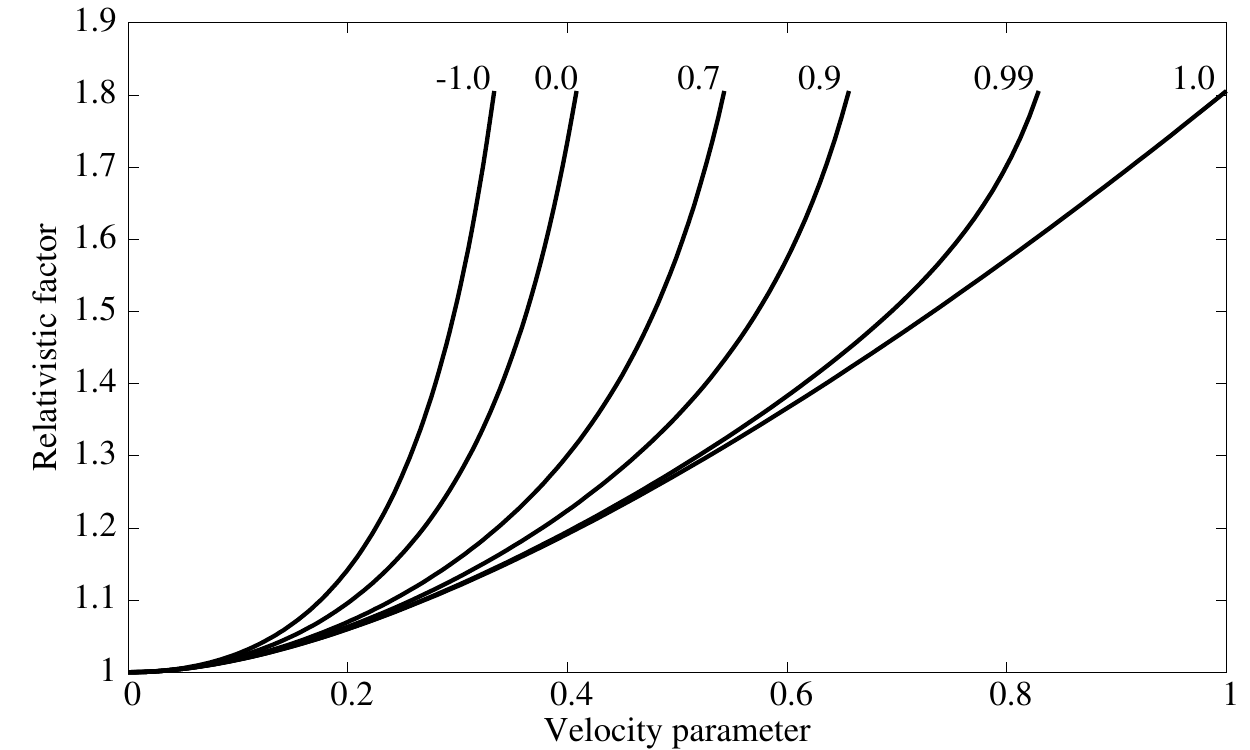}
\caption{Relativistic factor $\Gamma_{\rm K}$ for a rapidly rotating
  black  hole, plotted as a function of the orbital-velocity parameter
  $V$. The curves were computed for $\chi = 1$, which indicates that 
  the small black hole is extremal. They are labeled by $\Xi$, 
  which ranges from $\Xi = -1$ (extremal Kerr black hole,
  counter-rotating orbit) to $\Xi = +1$ (extremal Kerr black hole,
  corotating orbit). For $\chi=1$ the unique limiting value of 
  $\Gamma_{\rm K}$ is $\frac{65}{36} \simeq 1.8056$.}     
\end{figure} 

The relativistic factors are plotted in Figs.~1 and 2 for selected
values of $\Xi$, the Kerr parameter of the large black hole. For each
curve the velocity parameter $V$ ranges from 0 to the maximum value  
$V_{\rm isco}$ that  corresponds to the innermost stable circular
orbit. We notice the interesting phenomenon that each relativistic
factor takes on a unique limiting value when 
$V \to V_{\rm isco}$. The limiting values are  
\begin{equation} 
\Gamma_{\rm S} \to \frac{20}{9}, \qquad 
\Gamma_{\rm K} \to \frac{5}{9} \frac{4 + 9\chi^2}{1+3\chi^2}. 
\label{isco} 
\end{equation} 
Making the substitutions in Eqs.~(\ref{Schw}) and (\ref{Kerr}) reveals
that in this limit, $dm/dv$ and $dj/dv$ can be expressed entirely
in terms of $V_{\rm isco}$, and no longer depend on $\Xi$. 

The scaling of $dm/dv$ as $(m/M)^6$ in the case of a nonrotating
black hole, and as $(m/M)^5$ in the case of a rapidly rotating black
hole, implies that the heating and torquing of a small black hole by
tidal fields produced by a much larger black hole are always
insignificant. The effect will never modify the motion of the small
black hole in a substantial manner, in spite of the fact that the
motion takes place in the deep relativistic field of the large black
hole. This circumstance might have produced large values for the
relativistic factors $\Gamma_{\rm S}$ and $\Gamma_{\rm K}$, but our
calculations show instead that they are bounded by the numbers
displayed in Eqs.~(\ref{isco}). The interest of our work, therefore,
is not in the fact that the effect might lead to interesting
observational consequences. It does not. It is instead in the fact
that in the small-hole approximation, the tidal heating and torquing
of the small black hole can be {\it computed to all orders in the
velocity parameter} $V$, which is permitted to approach the speed of
light; and the exercise reveals that relativistic effects cannot
compensate for the smallness of $m/M$.     

For the tidal heating and torquing of the black hole to be
significant, it is necessary to increase the value of the mass
ratio. This can be done within the small-hole/slow-motion
approximation at the expense of reducing the orbital velocity
so that $V \ll 1$. This regime was examined by Taylor and Poisson
\cite{taylor-poisson:08}, who show that in this case the dependence of
$dm/dv$ upon the masses becomes 
\begin{equation} 
\biggl( \frac{m}{M} \biggr)^6 \to 
\frac{m^6 M^2}{(m+M)^8} 
\end{equation} 
when the black hole is nonrotating, and 
\begin{equation} 
\biggl( \frac{m}{M} \biggr)^5 \to 
\frac{m^5 M^2}{(m+M)^7} 
\end{equation} 
when the black hole is rapidly rotating. In the nonrotating case the 
mass-dependent factor is maximized when $m=3M$ and is then numerically 
equal to $0.01112$. In the rapidly-rotating case it is maximized when 
$m=\frac{5}{2}M$ and is then equal to $0.01483$. For such mass ratios
the effect is likely to be significant when $V$ becomes comparable to
unity. In the regime where $m$ is much larger than $M$, $dm/dv$
scales as $(M/m)^2$ in each case, and this is the same scaling that is
found for the rate at which energy is radiated away by gravitational
waves; this was the regime examined by Hughes \cite{hughes:01} and
Martel \cite{martel:04}, where the effect was shown to be
significant. 

This concludes the presentation of our results and the discussion of
their significance. In the following section of the paper we describe
how the results displayed in Eqs.~(\ref{Schw}), (\ref{Kerr}),
and (\ref{isco}) were obtained.   

\section{Derivations}  

For the purposes of computing tidal fields, the small black hole can
be thought of as a test particle moving on a circular orbit in the
equatorial plane of the Kerr spacetime. The large black hole has a
mass $M$ and angular momentum $J = \Xi M^2$, and the orbital radius is
$r$ in Boyer-Lindquist coordinates; the orbit is corotating with the
black hole when $\Xi > 0$, and it is counter-rotating when $\Xi <
0$. In place of $r$ it is useful to introduce the velocity parameter 
$V := \sqrt{M/r}$ to describe the orbit. In Boyer-Lindquist
coordinates $(t,r,\theta,\phi)$ the velocity vector of the small
black hole is $u^\alpha = \gamma(1,0,0,\Omega)$, with
\cite{bardeen-etal:72} 
\begin{equation} 
\Omega = \frac{\sign{\Xi}}{M} \frac{V^3}{1 + \Xi V^3}, \qquad 
\gamma = \frac{1+\Xi V^3}{\sqrt{1 - 3V^2 + 2\Xi V^3}};  
\end{equation} 
$\Omega$ is the orbital angular velocity of the small black hole. The
orbital velocity $V$ is limited to the interval $0 < V \leq 
V_{\rm isco}$, where $V_{\rm isco}$ is a solution to the quartic
equation \cite{bardeen-etal:72} 
\begin{equation} 
1 - 6V^2 + 8\Xi V^3 - 3\Xi^2 V^4 = 0. 
\label{quartic} 
\end{equation} 
This equation can also be viewed as a quadratic equation for
$\Xi$, whose solution gives $\Xi$ expressed as a function of 
$V_{\rm isco}$; this interpretation is useful for our purposes below.  

The tidal fields created by the large black hole are represented by   
components of the Weyl tensor evaluated as functions of proper time
$\tau$ on the world line of the small black hole. In terms of a tetrad
$(u^\alpha,e^\alpha_a)$ of orthonormal vectors that are
parallel-transported on the world line (with the lower-case Latin
index $a = \{1,2,3\}$ representing a spatial label), we have the
electric-type components $\E_{ab} := 
C_{\mu\alpha\nu\beta} u^\mu e^\alpha_a u^\nu e^\beta_b$ and the
magnetic-type components $\B_{ab} := \frac{1}{2} u^\mu e^\alpha_a  
\varepsilon_{\mu\alpha\gamma\delta} C^{\gamma\delta}_{\ \ \beta\nu}
e^\beta_b u^\nu$, where $C_{\mu\alpha\nu\beta}$ is the Weyl
tensor and
$\varepsilon_{\mu\alpha\gamma\delta}$ the Levi-Civita tensor. The
tidal moments $\E_{ab}$ and $\B_{ab}$ are symmetric and tracefree, in
the sense that $\E_{ba} = \E_{ab}$ and $\delta^{ab} \E_{ab} = 0$ (with
similar equations holding for $\B_{ab}$). 

An orthonormal tetrad adapted to equatorial, circular orbits of a Kerr
black hole can be constructed by specializing the general work of
Marck \cite{marck:83} to this specific case. We find 
\begin{subequations} 
\begin{align} 
e^\alpha_1 &= (-\mu \sin\Phi, \lambda \cos\Phi, 0, -\nu \sin\Phi), \\  
e^\alpha_2 &= (\mu \cos\Phi, \lambda \sin\Phi, 0, \nu \cos\Phi), \\ 
e^\alpha_3 &= (0,0,-V^2/M,0), 
\end{align}
\end{subequations} 
where 
\begin{subequations} 
\begin{align} 
\lambda &= \sqrt{1-2V^2+\Xi^2 V^4}, \\ 
\mu &= \frac{ \sign{\Xi} V (1-2\Xi V^3 + \Xi^2 V^4) }
  { \sqrt{1-2V^2+\Xi^2 V^4} \sqrt{1-3V^2+2\Xi V^3} }, \\ 
\nu &= \frac{ V^2 (1 - 2V^2 + \Xi V^3) }
  {M \sqrt{1-2V^2+\Xi^2 V^4} \sqrt{1-3V^2+2\Xi V^3} }, 
\end{align} 
\end{subequations} 
and 
\begin{equation} 
\Phi = \Omega^\dagger \tau, \qquad 
\Omega^\dagger =  \frac{\sign{\Xi}}{M} V^3. 
\end{equation} 
Notice that the basis is right-handed, and that $e^\alpha_3$ points in 
the direction of the angular-momentum vector of the large black
hole. The angular frequency of the tetrad vectors is $\Omega^\dagger$,
and this differs from the orbital angular velocity $\Omega$ because
(i) $\Omega^\dagger$ refers to proper time $\tau$ instead of
coordinate time $t$, and (ii) the tetrad vectors undergo geodetic and
Lense-Thirring precession relative to the global inertial frame.   

The nonvanishing components of the tidal fields $\E_{ab}$ and
$\B_{ab}$ are  
\begin{subequations} 
\allowdisplaybreaks
\label{EBfields} 
\begin{align} 
& \frac{1}{2}( \E_{11}+\E_{22} ) = -\frac{V^6}{2M^2} 
\frac{ 1 - 4\Xi V^3 + 3 \Xi^2 V^4 }{ 1 - 3V^2 + 2\Xi V^3 }
= -\frac{1}{2} \E_{33}, \\ 
& \frac{1}{2} ( \E_{11}-\E_{22} ) = -\frac{3V^6}{2M^2} 
\frac{ 1 - 2V^2 + \Xi^2 V^4 }{ 1 - 3V^2 + 2\Xi V^3 }
\cos 2\Phi, \\ 
& \E_{12} = -\frac{3V^6}{2M^2} 
\frac{ 1 - 2V^2 + \Xi^2 V^4 }{ 1 - 3V^2 + 2\Xi V^3 }
\sin 2\Phi, \\ 
& \B_{13} = -\frac{3\sign{\Xi} V^7}{M^2} 
\frac{(1-\Xi V)\sqrt{1-2V^2+\Xi^2 V^4}}{1-3V^2+2\Xi V^3} 
\cos\Phi, \\ 
& \B_{23} = -\frac{3\sign{\Xi} V^7}{M^2} 
\frac{(1-\Xi V)\sqrt{1-2V^2+\Xi^2 V^4}}{1-3V^2+2\Xi V^3} 
\sin\Phi. 
\end{align}
\end{subequations}
These equations generalize Eqs.~(2.107) and (2.108) of
Ref.~\cite{poisson:04a}, which hold for $\Xi = 0$ only. 

The limiting values of the tidal fields when $V \to V_{\rm isco}$ can
be simplified by solving Eq.~(\ref{quartic}) for 
$\Xi(V_{\rm isco})$. Making the substitution reveals that 
$\frac{1}{2}(\E_{11} +\E_{22}) \to -M^{-2} V_{\rm isco}^6$, 
$\frac{1}{2}(\E_{11}-\E_{22}) \to -2 M^{-2} V_{\rm isco}^6\cos 2\Phi$,  
$\E_{12} \to  -2 M^{-2} V_{\rm isco}^6\sin 2\Phi$, 
$\B_{13} \to -2\sign{\Xi} M^{-2} V_{\rm isco}^6\cos\Phi$, and 
$\B_{23} \to -2\sign{\Xi} M^{-2} V_{\rm isco}^6\sin\Phi$. 

According to Eqs.~(8.38) and (8.39) of Ref.~\cite{poisson:04d}, the
rates at which the mass and angular momentum of a nonrotating black
hole change as a result of a tidal interaction are given by 
$dm/dv = \frac{16}{45} m^6 ( \dot{\cal E}_{ab} \dot{\cal E}^{ab} 
+ \dot{\cal B}_{ab} \dot{\cal B}^{ab})$ and  
$dj/dv = -\frac{32}{45} m^6 \varepsilon_{acd} ( 
\dot{\cal E}^a_{\ b} {\cal E}^{bc} + \dot{\cal B}^a_{\ b} 
{\cal B}^{bc} ) s^d$,  
where an overdot indicates differentiation with respect to $\tau$, 
$\varepsilon_{acd}$ is the three-dimensional permutation symbol, and
$s^a = (0,0,1)$ denotes the direction of the angular momentum of the
large black hole. Substitution of the tidal fields of
Eqs.~(\ref{EBfields}) gives rise to Eqs.~(\ref{Schw}).  

On the other hand, Eq.~(9.39) of Ref.~\cite{poisson:04d} reveals that
the rate at which the angular momentum of a rapidly rotating black
hole changes as a result of a tidal interaction is given by 
$dj/dv = -\frac{2}{45} m^5\chi [ 8(1 + 3\chi^2) (E_1 + B_1 )  
- 3(4 + 17\chi^2) (E_2 + B_2 ) + 15\chi^2 ( E_3 + B_3 ) ]$,  
where $E_1 := \E_{ab} \E^{ab}$, $E_2 := (\E_{ab} s^b)(\E^a_{\ c}
s^c)$, $E_3 := (\E_{ab} s^a s^b)^2$, and $B_1 := \B_{ab} \B^{ab}$, 
$B_2 := (\B_{ab} s^b)(\B^a_{\ c}s^c)$, $B_3 := (\B_{ab} s^a
s^b)^2$. When the black hole is a member of a system in rigid
rotation, we also have $dm/dv = \Omega^\dagger dj/dv$.   
Substitution of the tidal fields of Eqs.~(\ref{EBfields}) gives rise
to Eqs.~(\ref{Kerr}).  

The limiting values of the relativistic factors $\Gamma_{\rm S}$ and
$\Gamma_{\rm K}$ when $V \to V_{\rm isco}$ can be simplified by
solving Eq.~(\ref{quartic}) for $\Xi(V_{\rm isco})$. When the
substitution is made into Eqs.~(\ref{GammaS}) and (\ref{GammaK}) we
find the results displayed in Eq.~(\ref{isco}). 

\begin{acknowledgments} 
This work was supported by the Natural Sciences and Engineering
Research Council of Canada.       
\end{acknowledgments} 

\bibliography{../bib/master}
\end{document}